# One-step synthesis of $K_xFe_{2-y}Se_2$ single crystal for high critical current density

Toshinori Ozaki,[1,3] Hiroyuki Takeya,[1,3] Hiroyuki Okazaki,[1,3] Keita Deguchi,[1,2,3] Satoshi Demura,[1,2,3] Yasuna Kawasaki,[1,2,3] Hiroshi Hara,[1,3] Tohru Watanabe,[1,2] Takahide Yamaguchi,[1,3] and Yoshihiko Takano[1,2,3]

[1] *National Institute for Materials Science, 1-2-1 Sengen, Tsukuba, Ibaraki 305-0047, Japan*

[2] *University of Tsukuba, 1-1-1Tennodai, Tsukuba, Ibaraki 305-0001, Japan*

[3] *JST, Transformative Research-Project on Iron Pnictides (TRIP), Chiyoda, Tokyo 102-0075, Japan*

(Abstract)

We have established a simple process that allows for the one-step synthesis of $K_xFe_{2-y}Se_2$ single crystals, which exhibit high critical current density $J_c$. The post annealing and quenching technique has improved the homogeneity of as-grown crystals, resulting in full shielding of the external magnetic field. The quenched crystals show a superconducting transition at $T_c^{onset}$ = 32.9 K and $T_c^{zero}$ = 32.1 K. The upper critical fields $\mu_0H_{c2}(0)$ for $H//ab$ and $H//c$ are estimated to be ~206 and ~50 T, respectively. The critical current densities $J_c$ for $H//ab$ and $H//c$ reach as high as $1.0\times10^5$ and $3.4\times10^4$ A/cm$^2$ at 5 K. Furthermore, $J_c$ exhibits a high field performance and a significantly weak temperature dependence up to 5 T, suggesting strong pinning. These results demonstrate that $K_xFe_{2-y}Se_2$ would be a promising candidate material for practical applications.

The recent discovery of superconductivity in iron selenide compounds $A_xFe_{2-y}Se_2$ (A = K, Cs, Rb, Tl/Rb, Tl/K)[1-6] with transition temperatures $T_c$ around 30 K has renewed interest in iron-based superconductors. Compared to other iron-based superconductors, $A_xFe_{2-y}Se_2$ exhibits several distinctive features: Fe deficiencies in the FeSe layer, proximity to an antiferromagnetic semiconducting state,[6] an extremely high Néel transition temperature[7], and the absence of hole pockets[8]. Interestingly, the interplay between superconductivity and magnetism can be tuned by changing the Fe-vacancy order.[9-14]

The high $T_c$, and high upper critical fields $H_{c2}$ as well as lower toxicity of these compounds compared to iron arsenides[1,2,15] will encourage their use in practical applications. Moreover, the post-annealing and quenching process is known to enhance the superconducting volume fraction and critical current density $J_c$ in $K_xFe_{2-y}Se_2$ single crystals.[14,16,17] To date, single crystals of $K_xFe_{2-y}Se_2$ have typically been grown using the Bridgman method and a self-flux method. However, these processes/synthesis are performed at high temperatures above the melting point of the reactants and through a long growth process. Here, we report the successful growth of $K_xFe_{2-y}Se_2$ single crystals with high $J_c$ using a simple one-step process at lower temperatures.

Single crystals of nominal composition $K_{0.8}Fe_2Se_2$ were grown as follows: Fe (99.9%), $K_2Se$ (99%) powders and Se grains (99.999%) were put into an alumina crucible and sealed into an evacuated quartz tube. The quartz tube was heated to 900°C and was kept constant for 12 hours. Following this, the tube was cooled down to room temperature by shutting down the furnace. As-grown crystals were sealed into an evacuated quartz tube and annealed at 400°C for 1 hour, followed by quenching in air as previously reported.[13]

The obtained crystals were characterized by x-ray diffraction with Cu-Ka radiation using the $2\theta$-$\omega$. The actual crystal composition was determined by energy dispersive x-ray (EDX) spectrometer. Resistivity measurements were carried out with a standard four-probe method in a Physical Property Measurement System (PPMS, Quantum Design). The magnetization was measured using a superconducting quantum interference device (SQUID, Quantum Design) magnetometer.

The as-grown single crystals show a dark shiny flat surface, as displayed in the inset of Fig 1(a). Figure 1(a) and 1(b) show the x-ray diffraction patterns for the as-grown and quenched $K_xFe_{2-y}Se_2$ single crystals. These two patterns are dominated by the (00*l*) diffraction; this confirms that the crystallographic *c* axis is perpendicular to the shiny surface. The peaks can be indexed to the *I*4/*mmm* space group. The lattice parameter *c* was estimated to be 14.1198(3) and 14.1087(16) Å for as-grown and quenched crystals, respectively. These parameters are consistent with previous reports.[1,2,18] We note that there is another series of (00*l*) peaks (marked by the asterisk) for the as-grown $K_xFe_{2-y}Se_2$ crystal, which are located at lower angles with respect to the main peaks. These weak peaks could arise from the inhomogeneous distribution of K atoms. Interestingly, the quenched crystal did not exhibit any sign of an additional series of (00*l*) peaks. This implies that the inhomogeneity of $K_xFe_{2-y}Se_2$ crystal was greatly decreased by the post-annealing and quenching process. The actual compositions of the single crystals were determined to be K:Fe:Se = 0.78:1.68:2 using an average of 5 points from the EDX measurements. This is indicative of deficiencies on both K and Fe sites.

Figure 2 shows the temperature dependence of magnetization for the as-grown and quenched $K_xFe_{2-y}Se_2$ single crystals with *H*//*c*. The superconducting transition temperatures $T_c^{mag}$ were estimated to be 31.0 and 31.2 K for the as-grown and quenched crystals, respectively. The quenching process had little effect on $T_c^{mag}$. However, the quenched crystal shows a very steep transition under zero-field-cooling (ZFC) condition with nearly full shielding, whereas the as-grown crystal exhibits a gradual transition that may be due to an inhomogeneity. This is consistent with existing reports.[10,16] These results indicate that the post-annealing and quenching process significantly increase the superconducting shielding volume fraction and homogeneity in the $K_xFe_{2-y}Se_2$ crystal.

Figure 3 displays the temperature dependence of the resistivity of the quenched $K_xFe_{2-y}Se_2$ single crystal at zero field, for temperatures ranging from 2 to 300 K, for a current along the *ab* plane. The quenched crystal exhibits a small resistivity and a minute hump around 250 K, which could be related to the order of the Fe vacancies. Below 250 K, the resistivity shows metallic behavior, and the crystal undergoes a very sharp superconducting transition around 33 K. The onset and zero-resistivity temperatures were estimated to be $T_c^{onset}$ = 32.9 K and $T_c^{zero}$ = 32.1 K. The obtained $T_c^{zero}$ is one of the highest values for $A_xFe_{2-y}Se_2$ single crystals.

The resistivity of the quenched $K_xFe_{2-y}Se_2$ single crystal as a function of temperature, under magnetic fields up to 7 T applied along the *c* (*H*//*c*) axis and within the *ab* plane (*H*//*ab*) are shown in Figs. 4(a) and 4(b). The superconducting transition is suppressed for both *H*//*c* and *H*//*ab*. The suppression is much larger when the field is applied along the *c* axis of the single crystals instead of the *ab* plane. This is indicative of strong vortex pinning within the *ab* plane for this configuration. We also found that the obtained $T_c$ values in each magnetic field are slightly higher than the reported values,[2,17,25] indicating that our crystal-growth method is well suited to grow $K_xFe_{2-y}Se_2$ single crystals. The temperature dependence of the upper critical fields $\mu_0H_{c2}$ was determined from the 90% and 10% resistivity drops right above the superconducting transition. The anisotropic $\mu_0H_{c2}$ is shown in Fig 4(c) for the two field directions. The $\mu_0H_{c2}$ curves show a linear temperature dependence for both orientations. The slopes $d\mu_0H_{c2}/dT|_{T=T_c}$of at $T_c^{90\%}$ and $T_c^{10\%}$ for both directions are listed in Table 1. The value of the slope of the two samples for the magnetic field with *ab* plane strongly exceeds the Pauli limit 1.84 T/K, which suggests an unconventional mechanism of superconductivity in this material. The zero-temperature upper critical field $\mu_0H_{c2}(0)$ can be calculated by using the Werthamer-Helfand-Hohenberg (WHH) model,[19] which gives $\mu_0H_{c2} = -0.693(d\mu_0H_{c2}/dT|_{T=T_c})T_c$ with the slope, $d\mu_0H_{c2}/dT$, determined from the 90% resistivity drop and $T_c = 32.7$. The values of $\mu_0H_{c2}(0)$ for *H*//*ab* and *H*//*c* were estimated to be ~206 T and ~50 T, respectively. The anisotropy $\gamma = H_{c2}^{ab}(0)/H_{c2}^{\ c}(0)$ is about 4, which is slightly larger than those in previous reports.[2,17,18,20] The zero temperature coherence length $\xi(0)$ can be estimated using the Ginzburg-Landau formula $\mu_0H_{c2}(0) = \Phi_0/2\pi\xi^2(0)$, where $\Phi_0 = 2.07\times10^{-15}$ Wb (Table 1).

Figure 5(a) and 5(b) show the magnetization hysteresis (*M-H*) loop of the quenched crystal for *H*//*c* and *H*//*ab* for fields up to 5 T. We observed symmetric *M-H* loops for both directions, suggesting that the bulk pinning controls mainly the entry and exit of the magnetic flux.[21,22] No fishtail effect was observed up to 5 T. The field dependence of critical current density $J_c$ was estimated from the *M-H* loops using the critical-state Bean model.[23,24] For a rectangular prism-shaped crystal of dimension $c < a < b$, when the magnetic field is applied along the *c* axis, the in-plane critical current density $J_c^{ab}$ according to this model is given by $J_c = 20\Delta M/[a(1-a/3b)]$, where $\Delta M$ is the width of the magnetic hysteresis loop for increasing and decreasing fields. When the magnetic field is applied parallel to the *ab* plane, two vortex motions both parallel to and across the planes are involved in the Bean model. However, the single crystal in our measurement is a long slab of thickness *c*. Assuming $a$ » $(c/3)\cdot(J_c^{ab}/J_c^{c})$, we obtain $J_c^{c} \approx 20\Delta M/c$.[24] The $J_c$ obtained for $H//c$ and $H//ab$ is shown in Fig. 5(c) and 5(d), respectively. The values of $J_c^{ab}(0)$ and $J_c^{c}(0)$ reached as high as $1.0\times10^5$ and $3.4\times10^4$ A/cm$^2$ at 5 K. In both cases $J_c$ is higher than previously reported and in the case of $H//ab$ $J_c$ is about two orders of magnitude greater.[16,18,25,26] Remarkably, the $J_c$ values are nearly independent of the field for both directions and exhibit a large critical current density even at high temperatures. Indeed, for temperatures below 20 K, $J_c^{c}$ is much higher than $2.0\times10^4$ A/cm$^2$ even up to fields of 5 T. This result is indicative of the strong pinning in $K_xFe_{2-y}Se_2$ single crystals, which suppresses the thermally activated flux motions. In addition, the $J_c$ values for $H//ab$ are rather higher than that for $H//c$, consistent with the results for $\mu_0H_{c2}$ for $H//c$ and $H//ab$. The ratio of $J_c^{c}(0)/J_c^{ab}(0)$ is approximately 3. These high values indicate that the $K_xFe_{2-y}Se_2$ single crystal exhibits a high current carrying ability throughout the entire temperature/magnetic field range in our measurements.

In conclusion, we have successfully grown $K_xFe_{2-y}Se_2$ single crystals using a simple one-step synthesis. The post-annealing and quenching technique drastically improved homogeneity and the superconducting volume fraction in the single crystal. The ZFC magnetic susceptibility demonstrates that the crystal is fully diamagnetic. We also presented the anisotropic transport and magnetic properties of the quenched $K_xFe_{2-y}Se_2$ single crystal. The upper critical fields $\mu_0H_{c2}(0)$ for $H//ab$ and $H//c$ are estimated to be ~206 and ~50 T, respectively. The anisotropy parameter $\gamma$ is calculated to be ~4. Magnetic hysteresis measurements show that $J_c^{ab}(0)$ and $J_c^{ab}(0)$ of the quenched $K_xFe_{2-y}Se_2$ single crystal can be as high as ~$1.0\times10^5$ A/cm$^2$ and ~$3.4\times10^4$ A/cm$^2$, respectively. Furthermore, the measurements of $J_c$ suggest that the material exhibits a superior high field performance and rather weak temperature dependence. These results suggest that the one-step synthesis is suitable for growing $K_xFe_{2-y}Se_2$ single crystals with high $J_c$ and this material would have a remarkable potential for practical applications.

“Acknowledgments”

This work was supported in part by the Japan Society for the Promotion of Science (JSPS) through Grants-in-Aid for JSPS Fellows and ‘Funding program for World-Leading Innovative R&D on Science Technology (FIRST) Program’. This research was supported by Strategic International Collaborative Research Program (SICORP), Japan Science and Technology Agency.

**References**

[1] GUO J., JIN S., WANG G., WANG S., ZHU K., ZHOU T., HE M. and /CHEN X., *Phys. Rev. B*, **82** (2010) 180520(R).

[2] MIZUGUCHI Y., TAKEYA H., KAWASAKI Y., OZAKI T., TSUDA S. YAMAGUCHI T. and TAKANO Y., *Appl. Phys. Lett.,* **98** (2011) 042511.

[3] KRZTON-MAZIOPA A., SHERMADINI Z., POMJAKUSHINA E., POMJAKUSHINA V., BENDELE M., AMOTO A., KHASANOV R., LUETKENS H. and CONDER K., *J. Phys.: Condens. Matter.,* **23** (2011) 052203.

[4] WANG A. F., YING J. J., YAN Y. J., LIU R. H., LUO X. G., LI Z. Y., WANG X. F., ZHANG M., YE G. J., CHENG P., XIANG Z. J. and CHEN X. H., *Phys. Rev. B*, **83** (2011) 060512.

[5] WANG H. D., DONG C. H., LI Z. J., MAO Q. H., SHU S. S., FENG C. M., YUAN H. Q. and FANG M. H., *Europhys. Lett.*, **93** (2011) 47004.

[6] FANG M. H., WANG H. D., DONG C. H., LI Z. J., FENG C. M., CHEN J. and YUAN Q. H., *Europhys. Lett.,* **94** (2011) 27009.

[7] BAO W., HUANG Q., CHEN G. F., GREEN M. A., WANG D. M., HE J. B., WANG X. Q. and QIU Y., *Chinese Phys. Lett.*, **28** (2011) 086104.

[8] ZHANG Y., YANG L. X., XU M., YE Z. R., CHEN F., HE C., JIANG J., XIE B. P., YING J. J., WANG X. F., CHEN X. H., HU J. P. and FENG D. L., *Nature Mater.*, **10** (2011) 273.

[9] RICCI A., POCCIA N, JOSEPH B., ARRIGHETTI G., BARBA L., PLAISIER J., CAMPI G., MIZUGUCHI Y., TAKEYA H., TAKANO Y., SAINI N. L. and MIANOCONI A., Supercond. Sci. Technol., **24** (2011) 082002.

[10] WANG Z., SONG Y. J., SHI H. L., WANG Z. W., CHEN Z., TIAN H. F., CHEN G. F., GUO J. G., YANG H. X. and LI J. Q., *Phys. Rev. B*, **83** (2011) 140505(R).

[11] ZAVALIJ P., BAO W., WANG X. F., YING J. J., CHEN, X. H., WANG D. M., HE J. B. WANG X. Q.CHEN. G. F., P. –Y. HSIEH, HUANG Q. and GREEN M. A., *Phys. Rev. B*, **83** (2011) 132509.

[12] LI W., DING H., DENG P., CHANG K., SONG C., HE K., WANG L., MA X., HU J. –P., CHEN X. and XUE Q. -K., *Nature Phys. Advanced Online Publication*, DOI: 10.1038/NPHYS2155 (2011).

[13] HAN F., SHEN B., WANG Z. -Y. and WEN H. –H., arXiv:1103.1347 (2011).

[14] RYU H., LEI H., FRENKEL A. I. and PETROVIC C., , arXiv:1111.2597 (2011).

[15] MUN E. D.,ALTARAWNEH M. M., MIELKE C. H., ZAPF V. S., HU R., BUD'KO S. L. and CANFIELD P. C., *Phys. Rev. B*, **83** (2011) 100514(R).

[16] LEI H. and PETROVIC C., *Phys. Rev. B*, **84** (2011) 212502.

[17] LEI H. and PETROVIC C., arXiv:1110.5316 (2011).

[18] LEI H. and PETROVIC C., *Phys. Rev. B*, **83** (2011) 184504.

[19] WERTHAMER N. R., HELFAND E. and HOHENBERG P. C., *Phys. Rev.*, **147** (1966) 295.

[20] WANG D. M., HE J. B., XIA T. -L. and CHEN G. F., *Phys. Rev. B*, **83** (2011) 132502.

[21] BURLACHKOV L., KOSHELEV A. E. and VINOKUR V. M., *Phys. Rev. B*, **54** (1996) 6750.

[22] PISSAS M., MORAITAKIS E., STAMOPOULOS D., PAPAVASSILIOU G., PSYCHAIRS V. and KOUTANDOS S., *J. Supercond. Incorp. Novel Magn.*, **14** (2001) 615.

[23] BEAN C. P., *Phys. Lett. B*, **8** (1962) 250.

[24] GYORGY E. M., VAN DOVER R. B., JACKSON K. A., SCHNEERMEYER L. F. and WASZCZAK J. V., *Appl. Phys. Lett.*, **55** (1989) 283.

[25] Gao Z., Qi Y., Wang L., Yao C., Wang D., Zhang X. and MA Y., arXiv:1110.5316 (2011).

[26] LEI H. and PETROVIC C., *Phys. Rev. B*, **84** (2011) 052507.

Table 1 $d\mu_0H_{c2}/dT|_{T=T_c}$, derived $\mu_0H_{c2}$(T) at and coherence length $\xi$(T) of quenched $K_xFe_{2-y}Se_2$ single crystals.$_c$

| | $d\mu_0H_{c2}/dT\|_{T_c}$ (T/K) | | $d\mu_0H_{c2}(0)$ (T) | $\xi(0)$ (nm) |
|---|---|---|---|---|
| | $T_c^{90\%}$ | $T_c^{10\%}$ | | |
| *H*//*c* | 2.19(4) | 1.50(1) | 50 | 2.56 |
| *H*//*ab* | 9.10(36) | 4.94(7) | 206 | 1.26 |

(Captions)

Fig. 1 The x-ray diffraction patterns of (a) as-grown and (b) quenched $K_xFe_{2-y}Se_2$ single crystals. The inset shows a photograph of as-grown $K_xFe_{2-y}Se_2$ single crystals (length scale 1 mm).

Fig. 2 Temperature dependence of magnetic susceptibility for both zero-field cooling (ZFC) and field cooling (ZFC) processes at a magnetic field of $H$ = 10 Oe applied in the *ab* plane for as-grown and quenched $K_xFe_{2-y}Se_2$ single crystals.

Fig. 3 Temperature dependence of the *ab*-plane resistivity for the quenched $K_xFe_{2-y}Se_2$ single crystal. The inset enlarges resistivity curve around the superconducting transition.

Fig. 4 Resistivity curves for quenched $K_xFe_{2-y}Se_2$ single crystal under both (a) $H//c$ and (b) $H//ab$ configuration. (c) Temperature dependence of $\mu_0H_{c2}$ for the quenched $K_xFe_{2-y}Se_2$ single crystal determined from 90% and 10% of resistivity drop in both orientations.

Fig. 5 Magnetic hysteresis loops at different temperatures of the quenched $K_xFe_{2-y}Se_2$ single crystal in fields parallel to (a) *c* axis and (b) *ab* plane. Magnetic-field dependence of critical current density (c) $J_c^{ab}(\mu_0H)$ and (d) $J_c^{c}(\mu_0H)$ calculated using the critical-state Bean model.

# Figure 1

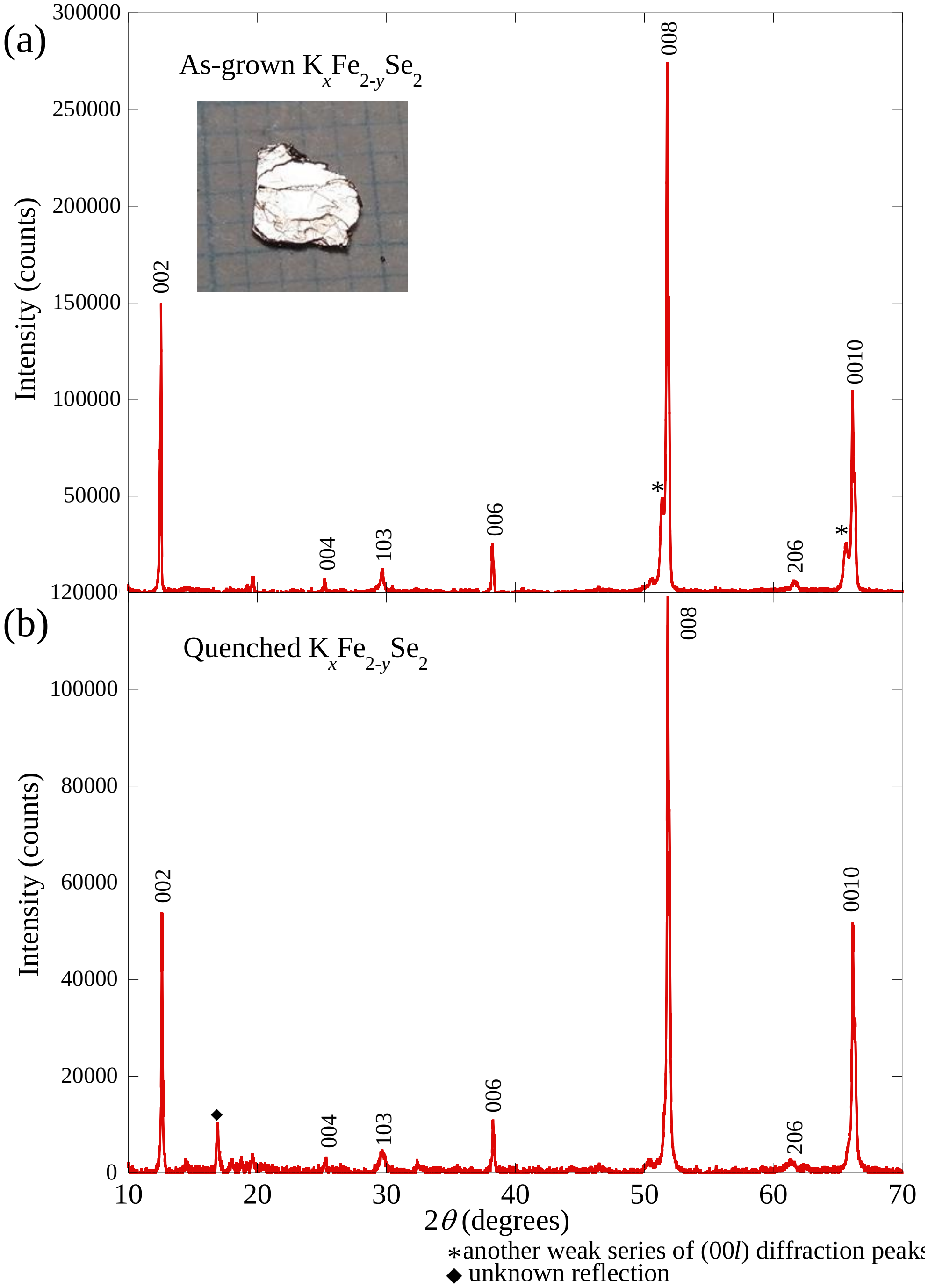

Figure 2

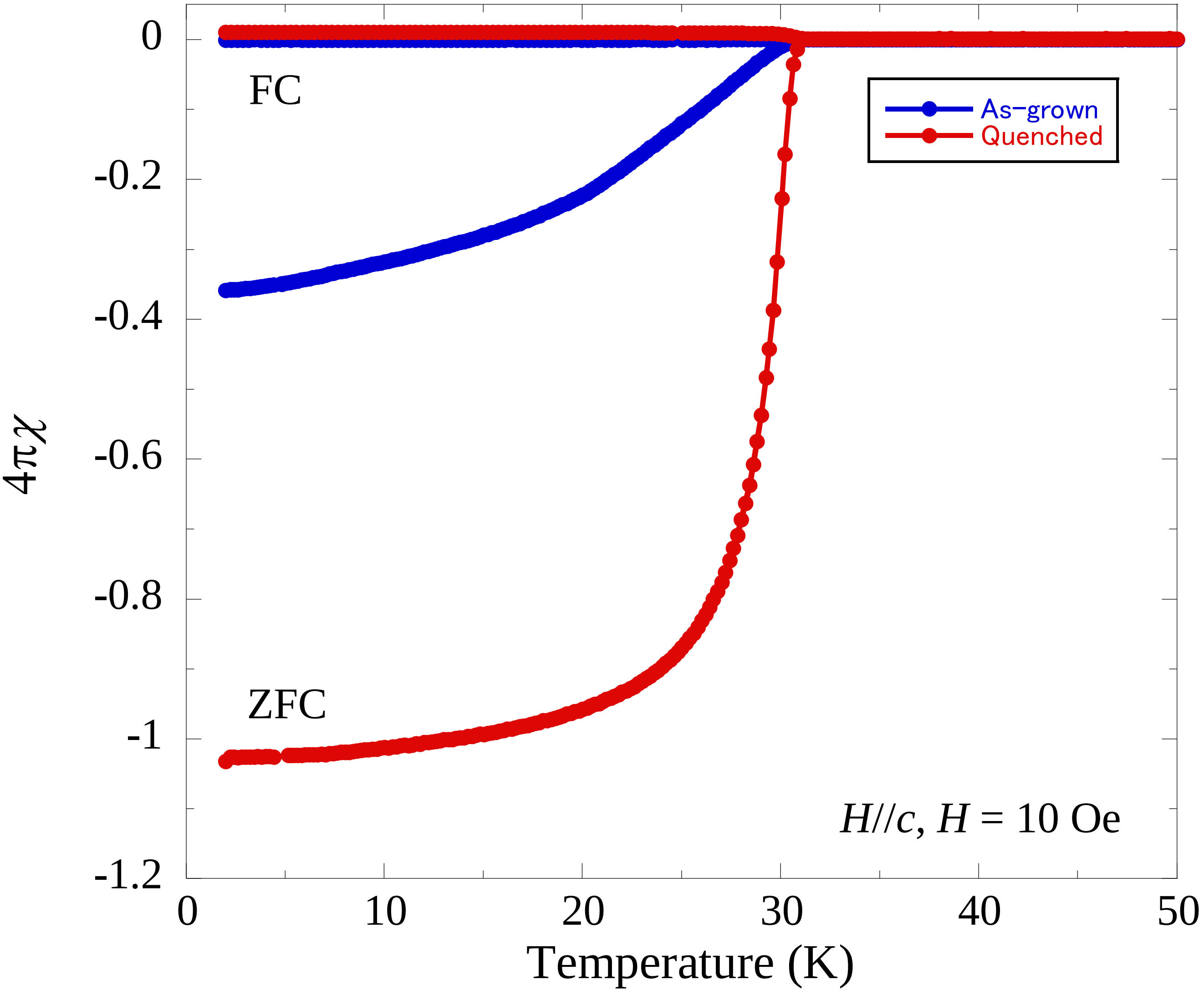

Figure 3

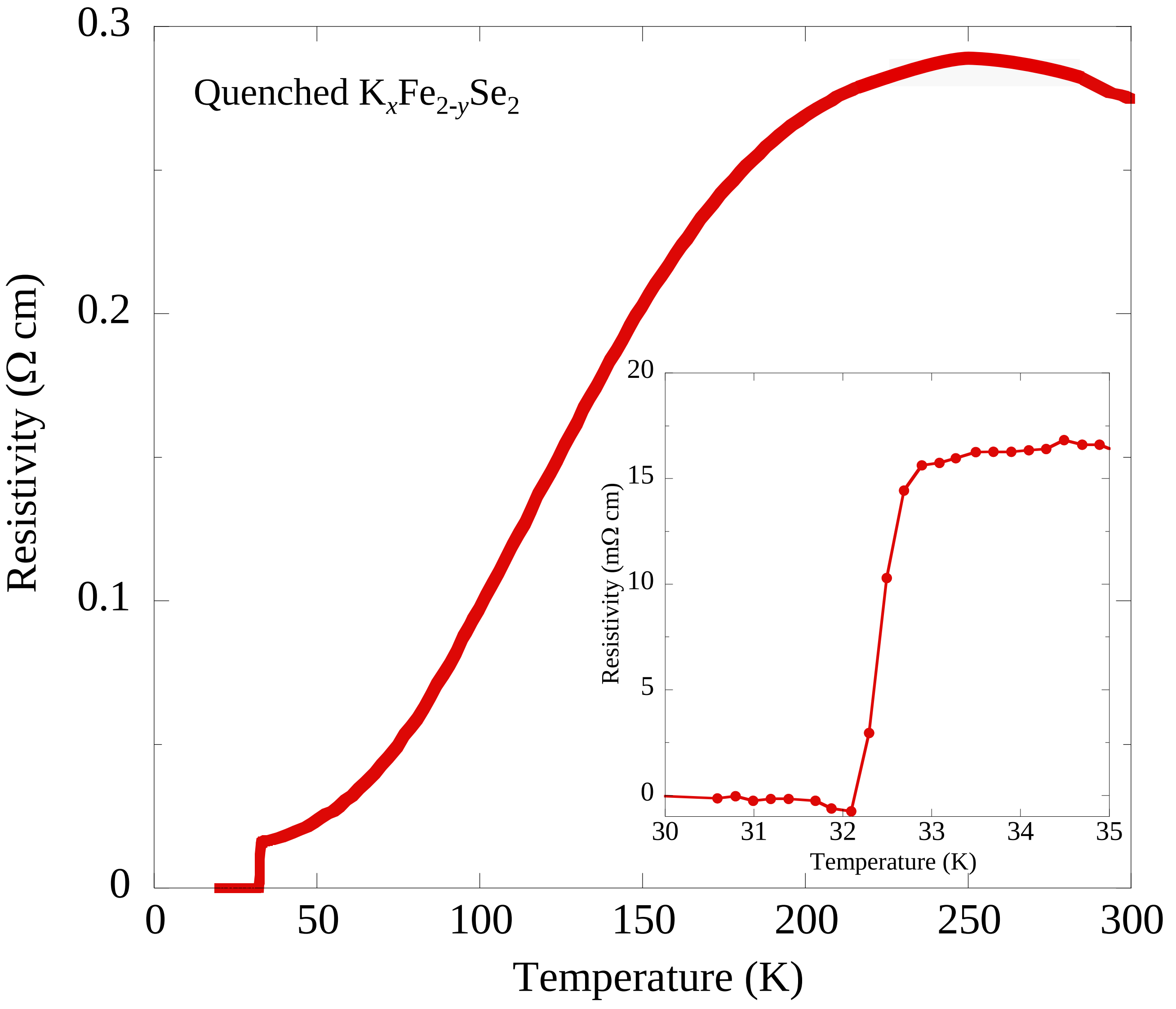

# Figure 4

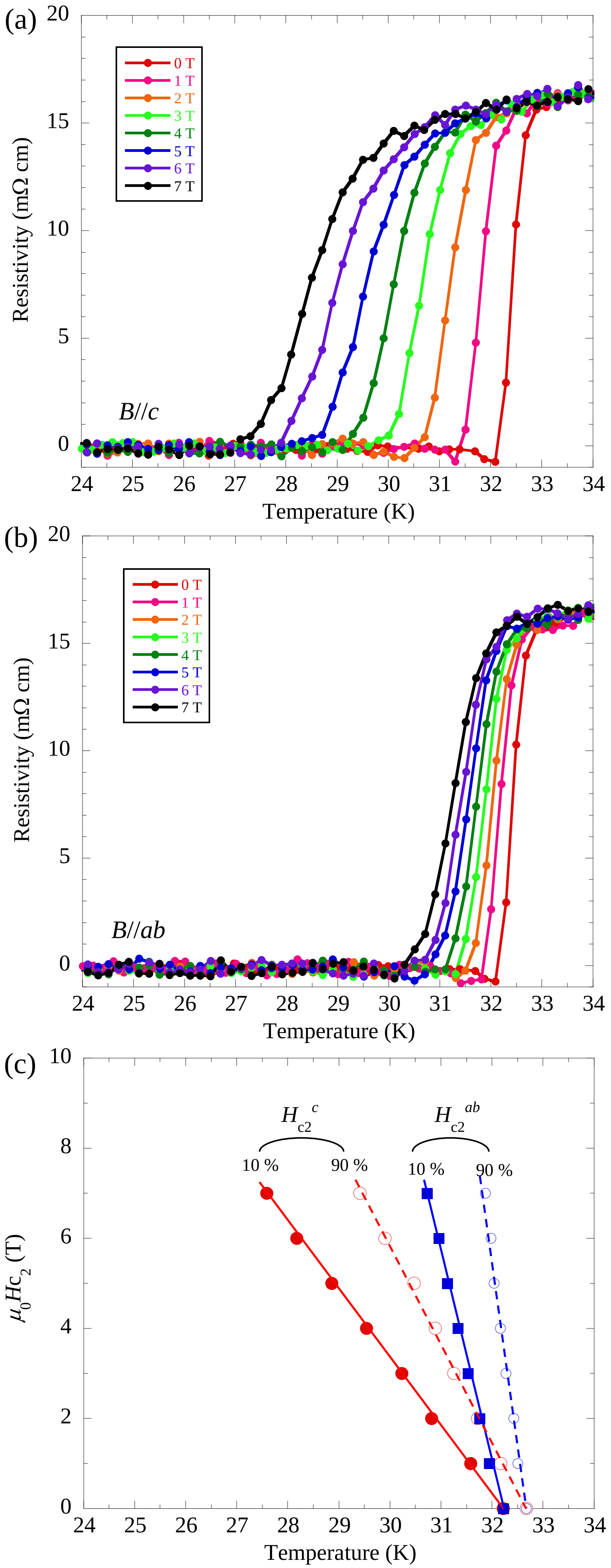

Figure 5

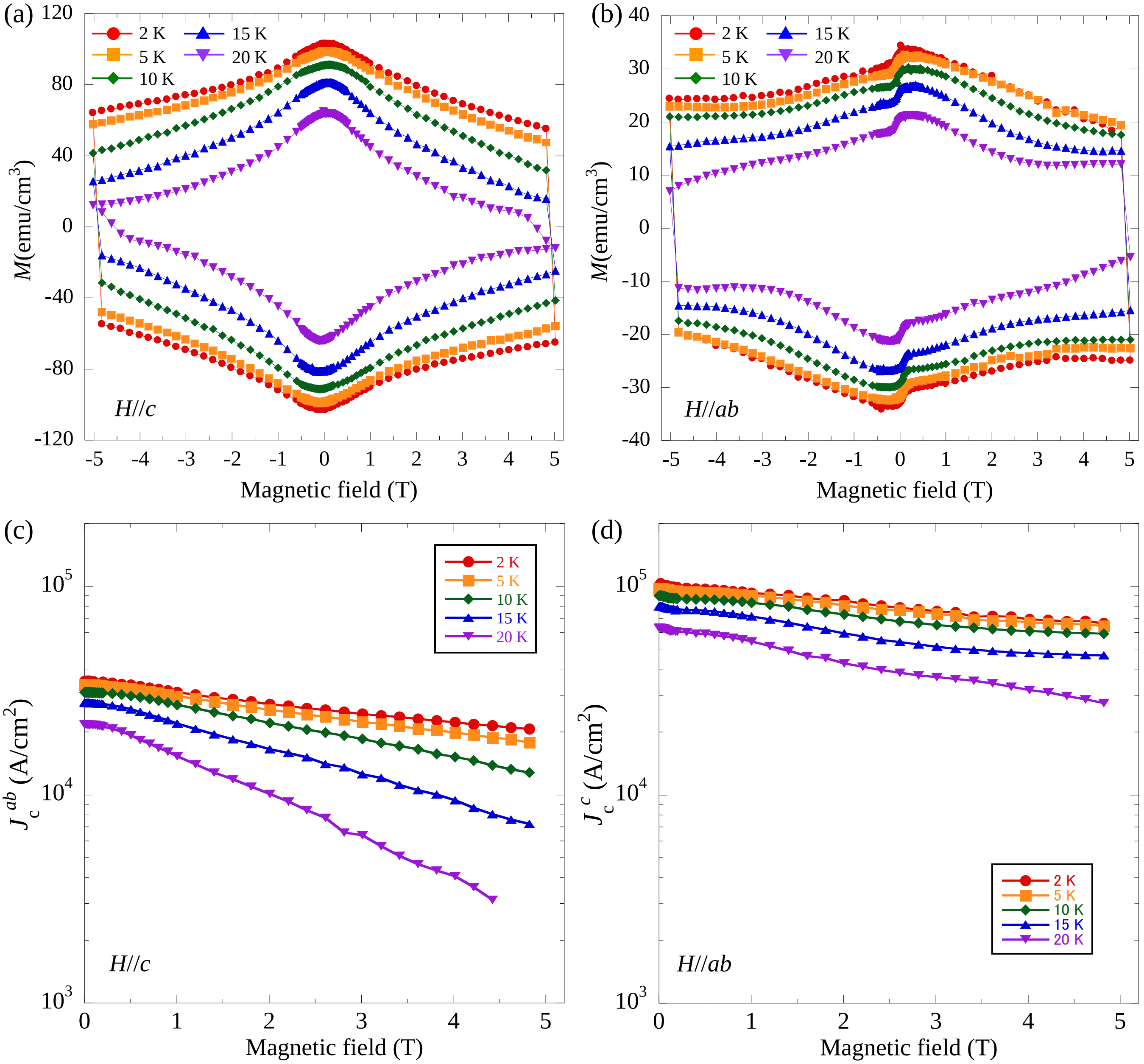